\documentclass[pdflatex]{sn-jnl} 
\usepackage{graphicx}%
\usepackage{multirow}%
\usepackage{amsmath,amssymb,amsfonts}%
\usepackage{amsthm}%
\usepackage{mathrsfs}%
\usepackage[title]{appendix}%
\usepackage{xcolor}%
\usepackage{color}%
\usepackage{textcomp}%
\usepackage{manyfoot}%
\usepackage{booktabs}%
\usepackage{algorithm}%
\usepackage{algorithmicx}%
\usepackage{algpseudocode}%
\usepackage{listings}%
\usepackage{comment}%
\usepackage{soul}
\usepackage{hyperref}
\usepackage{fullpage}
\usepackage{bm,txfonts}
\usepackage[modulo]{lineno}
\usepackage{array}
\usepackage[super]{natbib} 

\begin{document}

\title[]{Tomographic beta-gamma spectroscopy of nuclear beta decay}


\def\tdli{State Key Laboratory of Dark Matter Physics, Key Laboratory for Particle Astrophysics and Cosmology (MoE), Shanghai Key Laboratory for Particle Physics and Cosmology, Tsung-Dao Lee Institute \& School of Physics and Astronomy, Shanghai Jiao Tong University, Shanghai 201210, China}
\def\sjtuphys{State Key Laboratory of Dark Matter Physics, Key Laboratory for Particle Astrophysics and Cosmology (MoE), Shanghai Key Laboratory for Particle Physics and Cosmology, School of Physics and Astronomy, Shanghai Jiao Tong University, Shanghai 200240, China}
\def\newcorner{New Cornerstone Science Laboratory, Tsung-Dao Lee Institute, Shanghai Jiao Tong University, Shanghai 201210, China}
\def\MESJTU{School of Mechanical Engineering, Shanghai Jiao Tong University, Shanghai 200240, China}
\def\SPEIT{SJTU Paris Elite Institute of Technology, Shanghai Jiao Tong University, Shanghai 200240, China}
\def\SJTUSC{Shanghai Jiao Tong University Sichuan Research Institute, Chengdu 610213, China}

\def\BUAA{School of Physics, Beihang University, Beijing 102206, China}
\def\BUAACenter{Peng Huanwu Collaborative Center for Research and Education, Beihang University, Beijing 100191, China}
\def\BUAALab{International Research Center for Nuclei and Particles in the Cosmos \& Beijing Key Laboratory of Advanced Nuclear Materials and Physics, Beihang University, Beijing 100191, China}
\def\SCNT{Southern Center for Nuclear-Science Theory (SCNT), Institute of Modern Physics, Chinese Academy of Sciences, Huizhou 516000, China}

\def\USTClab{State Key Laboratory of Particle Detection and Electronics, University of Science and Technology of China, Hefei 230026, China}
\def\USTCdep{Department of Modern Physics, University of Science and Technology of China, Hefei 230026, China}

\def\YaLongSD{Yalong River Hydropower Development Company, Ltd., 288 Shuanglin Road, Chengdu 610051, China}
\def\scKeyLab{Jinping Deep Underground Frontier Science and Dark Matter Key Laboratory of Sichuan Province, Liangshan 615000, China}

\def\pku{School of Physics, Peking University, Beijing 100871, China}
\def\CHEPpku{Center for High Energy Physics, Peking University, Beijing 100871, China}

\def\SDUdep{Research Center for Particle Science and Technology, Institute of Frontier and Interdisciplinary Science, Shandong University, Qingdao 266237, China}
\def\SDUlab{Key Laboratory of Particle Physics and Particle Irradiation of Ministry of Education, Shandong University, Qingdao 266237, China}

\def\UMD{Department of Physics, University of Maryland, College Park, Maryland 20742, USA}

\def\SYU{School of Physics, Sun Yat-Sen University, Guangzhou 510275, China}
\def\SYUSFI{Sino-French Institute of Nuclear Engineering and Technology, Sun Yat-Sen University, Zhuhai 519082, China}
\def\SYUzhuhai{School of Physics and Astronomy, Sun Yat-Sen University, Zhuhai 519082, China}
\def\SYUshenzhen{School of Science, Sun Yat-Sen University, Shenzhen 518107, China}

\def\NKU{School of Physics, Nankai University, Tianjin 300071, China}
\def\YTU{Department of Physics, Yantai University, Yantai 264005, China}
\def\FDU{Key Laboratory of Nuclear Physics and Ion-beam Application (MOE), Institute of Modern Physics, Fudan University, Shanghai 200433, China}
\def\CDUT{College of Nuclear Technology and Automation Engineering, Chengdu University of Technology, Chengdu 610059, China}

\def\JYU{Department of Physics, University of Jyväskylä, P.O. Box 35, FI-40014 Jyväskylä, Finland}
\def\UOY{School of Physics, Engineering and Technology, University of York, Heslington, York YO10 5DD, United Kingdom}
\def\CIFRA{International Centre for Advanced Training and Research in Physics (CIFRA), P.O. Box MG12, 077125 Bucharest-Măgurele, Romania}

\author[2]{PandaX Collaboration: Zhe Yuan}
\author[3]{Zihao Bo}
\author[3]{Wei Chen}
\author[1,4,5]{Xun Chen}
\author[5,6]{Yunhua Chen}
\author[7]{Chen Cheng}
\author[1]{Xiangyi Cui}
\author[8]{Manna Deng}
\author[9]{Yingjie Fan}
\author[2]{Deqing Fang}
\author[3]{Xuanye Fu}
\author[3]{Zhixing Gao}
\author[8]{Yujie Ge}
\author[7,10,11,12]{Lisheng Geng}
\author[3,5]{Karl Giboni}
\author[7]{Xunan Guo}
\author[5,6]{Xuyuan Guo}
\author[7]{Zichao Guo}
\author[1]{Chencheng Han}
\author*[3,4,5]{Ke Han}\email{ke.han@sjtu.edu.cn}
\author[3]{Changda He}
\author[6]{Jinrong He}
\author[13]{Houqi Huang}
\author[3,5]{Junting Huang}
\author[3]{Yule Huang}
\author[4,5]{Ruquan Hou}
\author[14]{Xiangdong Ji}
\author[5,15]{Yonglin Ju}
\author[16]{Xiaorun Lan}
\author[3]{Chenxiang Li}
\author[17]{Jiafu Li}
\author[5,6]{Mingchuan Li}
\author[3]{Peiyuan Li}
\author[3,5,6]{Shuaijie Li}
\author[13]{Tao Li}
\author[3]{Yangdong Li}
\author[8]{Zhiyuan Li}
\author[16,18]{Qing Lin}
\author[1,4,5,19]{Jianglai Liu}\equalcont{Spokesperson}
\author[3]{Yuanchun Liu}
\author[15]{Congcong Lu}
\author[20,21]{Xiaoying Lu}
\author[22]{Lingyin Luo}
\author[16]{Yunyang Luo}
\author[2]{Yugang Ma}
\author[22]{Yajun Mao}
\author[3,4,5]{Yue Meng}
\author[20,21]{Binyu Pang}
\author[5,6]{Ningchun Qi}
\author[3]{Zhicheng Qian}
\author[20,21]{Xiangxiang Ren}
\author[23]{Dong Shan}
\author[3]{Xiaofeng Shang}
\author[23]{Xiyuan Shao}
\author[7]{Guofang Shen}
\author[5,6]{Manbin Shen}
\author[5,6]{Wenliang Sun}
\author[3]{Xuyan Sun}
\author[24]{Yi Tao}
\author[7]{Yueqiang Tian}
\author[3]{Yuxin Tian}
\author[20,21]{Anqing Wang}
\author[3]{Guanbo Wang}
\author[3]{Hao Wang}
\author[3]{Haoyu Wang}
\author[1]{Jiamin Wang}
\author[25]{Lei Wang}
\author[20,21]{Meng Wang}
\author[2]{Qiuhong Wang}
\author[3,5,13]{Shaobo Wang}
\author[15]{Shibo Wang}
\author[22]{Siguang Wang}
\author[8,17]{Wei Wang}
\author[1]{Xu Wang}
\author[1,4,5]{Zhou Wang}
\author[8]{Yuehuan Wei}
\author[3,5]{Weihao Wu}
\author[3]{Yuan Wu}
\author[3]{Mengjiao Xiao}
\author[17]{Xiang Xiao}
\author[5,6]{Kaizhi Xiong}
\author[3]{Jianqin Xu}
\author[15]{Yifan Xu}
\author[13]{Shunyu Yao}
\author[1]{Binbin Yan}
\author[26]{Xiyu Yan}
\author[3,5]{Yong Yang}
\author[3]{Peihua Ye}
\author[23]{Chunxu Yu}
\author[3]{Ying Yuan}
\author[3]{Youhui Yun}
\author[3]{Xinning Zeng}
\author[1]{Minzhen Zhang}
\author[5,6]{Peng Zhang}
\author[1]{Shibo Zhang}
\author[17]{Siyuan Zhang}
\author[17]{Shu Zhang}
\author[1,4,5]{Tao Zhang}
\author[1]{Wei Zhang}
\author[20,21]{Yang Zhang}
\author[20,21]{Yingxin Zhang}
\author[1]{Yuanyuan Zhang}
\author[1,4,5]{Li Zhao}
\author[1]{Kangkang Zhao}
\author[5,6]{Jifang Zhou}
\author[13]{Jiaxu Zhou}
\author[1]{Jiayi Zhou}
\author[1,4,5]{Ning Zhou}
\author[7]{Xiaopeng Zhou}
\author[3]{Zhizhen Zhou}
\author[16]{Chenhui Zhu}

\author[27,28]{\\\\Marlom Ramalho}
\author[27,29]{Jouni Suhonen}

\affil[1]{\tdli}
\affil[2]{\FDU}
\affil[3]{\sjtuphys}
\affil[4]{\SJTUSC}
\affil[5]{\scKeyLab}
\affil[6]{\YaLongSD}
\affil[7]{\BUAA}
\affil[8]{\SYUSFI}
\affil[9]{\YTU}
\affil[10]{\BUAACenter}
\affil[11]{\BUAALab}
\affil[12]{\SCNT}
\affil[13]{\SPEIT}
\affil[14]{\UMD}
\affil[15]{\MESJTU}
\affil[16]{\USTCdep}
\affil[17]{\SYU}
\affil[18]{\USTClab}
\affil[19]{\newcorner}
\affil[20]{\SDUdep}
\affil[21]{\SDUlab}
\affil[22]{\pku}
\affil[23]{\NKU}
\affil[24]{\SYUshenzhen}
\affil[25]{\CDUT}
\affil[26]{\SYUzhuhai}
\affil[27]{\JYU}
\affil[28]{\UOY}
\affil[29]{\CIFRA}
\abstract{
\textbf{Nuclear $\beta$ decay, a sensitive probe of nuclear structure and weak interactions, has become a precision test bed for physics beyond the Standard Model, driven by recent advances in spectrometric techniques.
Here we introduce tomographic $\beta$-$\gamma$ spectroscopy (TBGS) of nuclear $\beta$ decay, a method that detects the energies of $\beta$, $\gamma$, and internal conversion electrons while simultaneously reconstructing the energy deposition vertices.
Using the PandaX-4T detector operated as a TBGS, we obtain a precise and unbiased decay scheme of $^{214}$Pb, a key background isotope in searches for dark matter and Majorana neutrinos.
For the first time, transitions of $^{214}$Pb to both the ground and excited states of $^{214}$Bi are measured concurrently, revealing discrepancies in branching ratios of up to 4.7$\sigma$ relative to previous evaluations.
Combined with state-of-the-art theoretical spectral-shape calculations, these results establish a new benchmark for background modelling in rare-event searches and highlight the potential of TBGS as a versatile tool for fundamental physics and nuclear applications.
}
}
\maketitle

\section{Introduction}

Our understanding of nuclear $\beta$ decay has evolved alongside advances in atomic, nuclear, and particle physics.
From Rutherford’s discovery of $\beta$ rays, through Fermi’s theory of $\beta$ decay and the revelation of parity violation, to modern neutrino physics, $\beta$ decay has remained at the centre of fundamental research.
Today, it serves as a powerful tool to constrain neutrino wavepackets~\cite{Smolsky:2024uby}, test the unitarity of the CKM matrix~\cite{Towner:2010zz}, probe new electroweak couplings~\cite{Burkey:2022gpb}, and measure the absolute neutrino mass~\cite{KATRIN:2024cdt}.

At the forefront of this research lies spectroscopy—the measurement of $\beta$ spectra ($\beta$ spectroscopy) and the relative intensities of de-excitation $\gamma$ rays ($\gamma$ spectroscopy).
A wide range of $\beta$ spectroscopic techniques have been developed, including ion traps~\cite{Alcala:2025enm}, cryogenic calorimeters~\cite{Pagnanini:2024qmi}, electrostatic filters~\cite{KATRIN:2024cdt}, and cyclotron-resonance spectroscopy~\cite{Project8:2022hun}.
Advanced $\gamma$-ray spectrometers, such as GRETA/GRETINA~\cite{PASCHALIS201344} and AGATA~\cite{Akkoyun:2011ft}, provide high-resolution data for nuclear level schemes and structure.
$\gamma$ spectroscopy also finds broad applications in trace-isotope detection for environmental monitoring~\cite{IAEAgamma} and in low-radioactivity material screening~\cite{PandaX-4T:2021lbm}.
Total-absorption $\gamma$ spectroscopy (TAGS), which measures the total $\gamma$-ray energy with high efficiency and avoids the Pandemonium effect~\cite{Hardy:1977suw}, plays a crucial role in reactor antineutrino flux estimation and nuclear astrophysics~\cite{IGISOL:2015ifm}.
While $\beta$–$\gamma$ timing correlations have been used to improve $\gamma$-spectroscopy quality~\cite{Nishimura:2012mh}, achieving simultaneous precision measurements of both $\beta$ and $\gamma$ spectra remains a major experimental challenge.

We propose a tomographic $\beta$–$\gamma$ spectroscopy (TBGS) technique for $\beta$ decays, demonstrated using a liquid-xenon time projection chamber (TPC) in the PandaX-4T experiment.
With a radioactive sample introduced into the xenon volume, the detector measures the energies of all decay products (excluding neutrinos)—including $\beta$ particles, de-excitation $\gamma$ rays, and internal conversion electrons (ICEs)—while reconstructing their local energy-deposition vertices with millimetre precision.
Compared with scintillator-based TAGS, TBGS enables a direct and unbiased reconstruction of the full decay scheme through simultaneous $\beta$–$\gamma$–ICE detection, improved efficiency, and better energy resolution.
When combined with theoretical spectral-shape calculations, TBGS yields the $\beta$ spectra of all major decay branches using a single detector and dataset.

The prowess of TBGS is demonstrated through a precise measurement of the $^{214}$Pb $\beta$ decay scheme—one of the dominant intrinsic backgrounds in liquid-xenon detectors used for dark matter and neutrino studies~\cite{Haselschwardt:2020iey}.
$^{214}$Pb originates from the decay of $^{222}$Rn, which emanates from detector materials and surrounding components.
Despite decades of purification efforts reducing $^{222}$Rn concentrations to below 0.1 atoms per mole of xenon, its progeny $^{214}$Pb continues to dominate low-energy backgrounds in next-generation detectors.
During a dedicated $^{222}$Rn calibration campaign, PandaX-4T collected approximately half a million $^{214}$Pb decay events between May 27 and June 9, 2022.
Using state-of-the-art theoretical $\beta$-spectrum calculations, we performed iterative fits to extract the best-fit branching ratios (BRs) and spectral shapes.
The result represents the most precise determination of the $^{214}$Pb $\beta$-decay scheme to date and the first demonstration of TBGS as a precision tool for improving nuclear data.


\section{The challenge of $\beta$ decay spectroscopy}
 \begin{figure}[b]
  \centering
    \includegraphics[width=0.48\columnwidth]{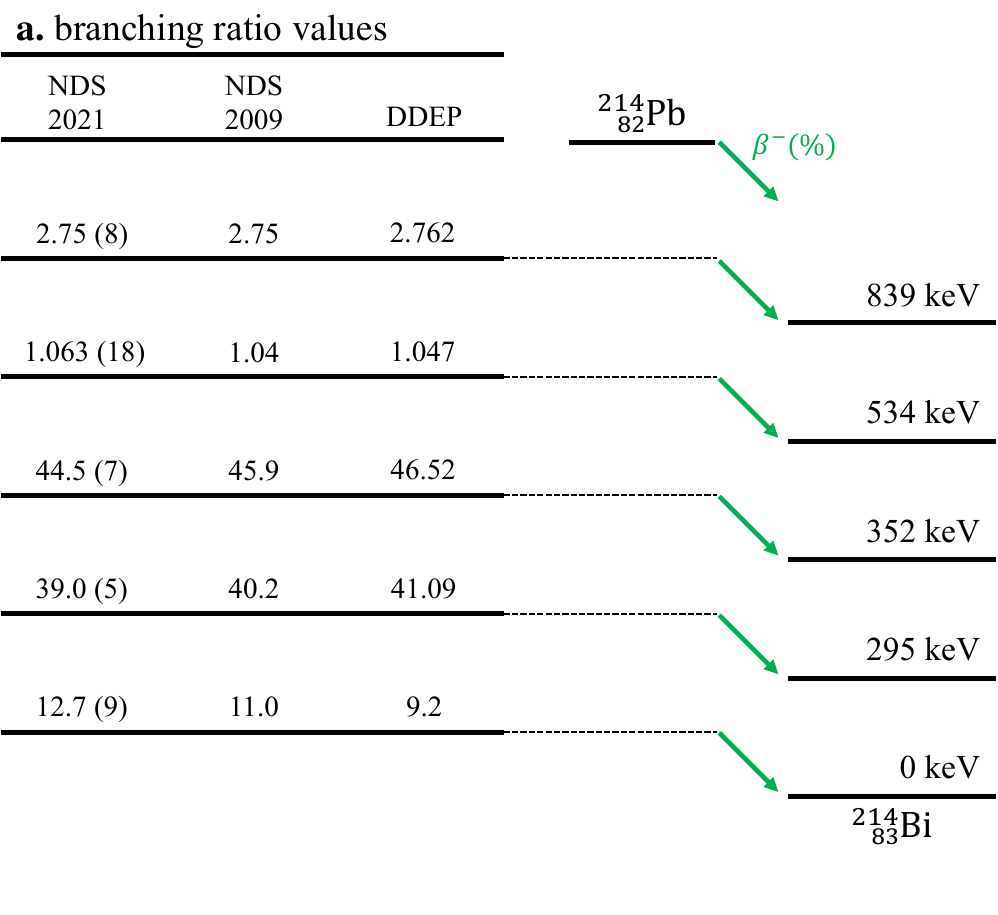}
    \hspace{0.05\columnwidth}
    \includegraphics[width=0.45\columnwidth]{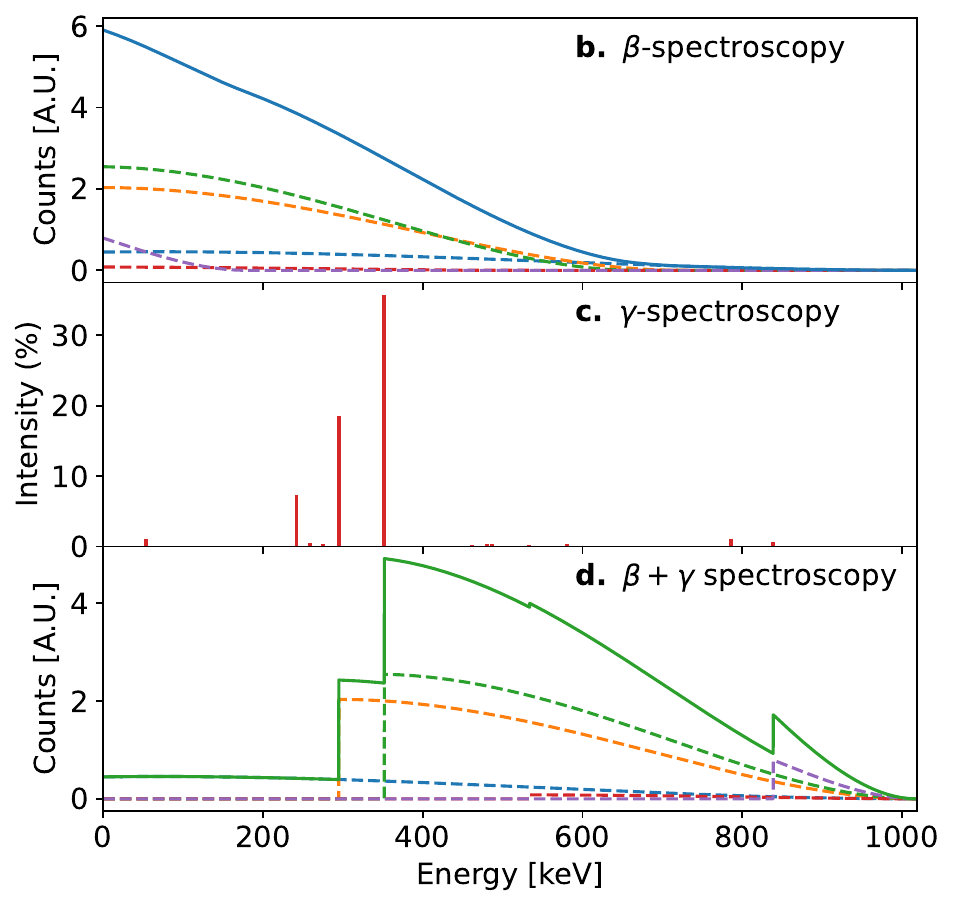}
    
    \caption{a. The $\beta$ decay scheme of $^{214}$Pb and the corresponding branching ratios in different databases. 
    Uncertainties follow the NDS2021 values in parentheses.
    No uncertainties are quoted in the other two databases. 
    b.c.d. Illustration of $\beta$ decay spectroscopic techniques.
    The spectra do not include detector effects.
    }
    \label{fig:Branching_ratio_database}
\end{figure}

$\beta$ decay (specifically $\beta^-$ decay) is a weak-interaction process in which a neutron in a nucleus transforms into a proton, emitting an electron (the $\beta$ particle) and an antineutrino.
The daughter nucleus may be in the ground state (GS) and various excited states (ESs), which subsequently de-excite through (cascades of) $\gamma$ rays and/or ICEs.
For example, Fig.~\ref{fig:Branching_ratio_database}a shows the major $^{214}$Pb decay branches to the GS and ESs of $^{214}$Bi, with a $Q$-value of 1019 keV.
We denote these branches according to their excitation energies as $\mathrm{GS}$, $\mathrm{ES_{295}}$, $\mathrm{ES_{352}}$, $\mathrm{ES_{534}}$, and $\mathrm{ES_{839}}$, respectively.

Traditional $\beta$ decay spectroscopy does not resolve the complete decay scheme.
The $\beta$ spectra from different branches overlap, all starting from 0 keV, making it nearly impossible to separate the individual components through conventional $\beta$ spectroscopy, as illustrated in Fig.~\ref{fig:Branching_ratio_database}b.
ICEs further distort the measured $\beta$ spectra.
Moreover, the relative $\gamma$ intensities obtained from $\gamma$ spectroscopy (Fig.~\ref{fig:Branching_ratio_database}c) do not directly represent the decay BRs.
To extract BRs, one must account for the internal conversion coefficient—the ratio of ICEs to $\gamma$ rays emitted in the de-excitation of each ES—and for the cascade de-excitation paths via intermediate states.
Additional energy release from subsequent decays of unstable daughter nuclei can further complicate the determination of BRs.

Spectroscopic detection of both $\beta$ and $\gamma$ emissions introduces additional uncertainties into the decay scheme.
Energy loss of $\beta$ particles along their trajectories, combined with energy-dependent detector efficiency and nonlinearity, inadvertently distorts the measured $\beta$ spectra.
Additional challenges arise in $\gamma$ spectroscopy, where relative intensity measurements depend critically on accurate modelling of detector response, typically based entirely on Monte Carlo (MC) simulations.

Figure~\ref{fig:Branching_ratio_database}a compares evaluated BRs from the Nuclear Data Sheets (NDS) published in 2021~\cite{Zhu:2021qss} and 2009~\cite{Wu:2009lpp}, as well as from the DDEP database, the de facto standard in metrology~\cite{DDEP}.
BRs for the ESs are derived from $\gamma$ intensity data and internal conversion coefficients~\cite{NNDC}, while the GS ratio is inferred such that the total probability sums to unity~\cite{Lingeman:intensity}.
Uncertainties in NDS2021 originate from the evaluation process~\cite{NNDC}, whereas the other two databases provide no quoted uncertainties.
The substantial discrepancies among these evaluations highlight the persistent challenges in $\beta$ decay spectroscopy.

Theoretical calculations of $\beta$ spectra are also highly nontrivial, particularly for forbidden transitions such as $\mathrm{GS}$, $\mathrm{ES_{295}}$, and $\mathrm{ES_{534}}$ in $^{214}$Pb.
The shape of each $\beta$ spectrum depends on both experimental inputs—such as end-point energies and BRs—and theoretical treatments of the nuclear structure of the initial and final states, the quenching of the axial-vector coupling constant $g_\textrm{A}$, atomic exchange effects, and other corrections.
These effects are treated differently in various theoretical models, leading to noticeable discrepancies in predicted spectral shapes~\cite{Haselschwardt:2020iey}.
A rigorous and consistent treatment of all branches, based on reliable experimental data, is therefore urgently needed.

The spectral shapes and BRs of $^{214}$Pb decays directly affect the physics reach of xenon-based detectors.
In dark matter (particularly Weakly Interacting Massive Particles, i.e., WIMP) and solar proton–proton neutrino searches, the most sensitive energy regions lie below 30 keV and 200 keV, respectively, where the $^{214}$Pb GS $\beta$ decay constitutes a dominant background.
At higher energies, the spectral shapes and BRs of ES decays introduce one of the leading systematic uncertainties in measurements of double-$\beta$ decay half-lives of xenon isotopes~\cite{PandaX:2022kwg, PandaX:2023ggs}, as well as in searches for new physics beyond the Standard Model~\cite{PandaX:2024sds}.

\section{PandaX-4T TPC for TBGS}
 \begin{figure}[t]
    \includegraphics[width=\columnwidth]{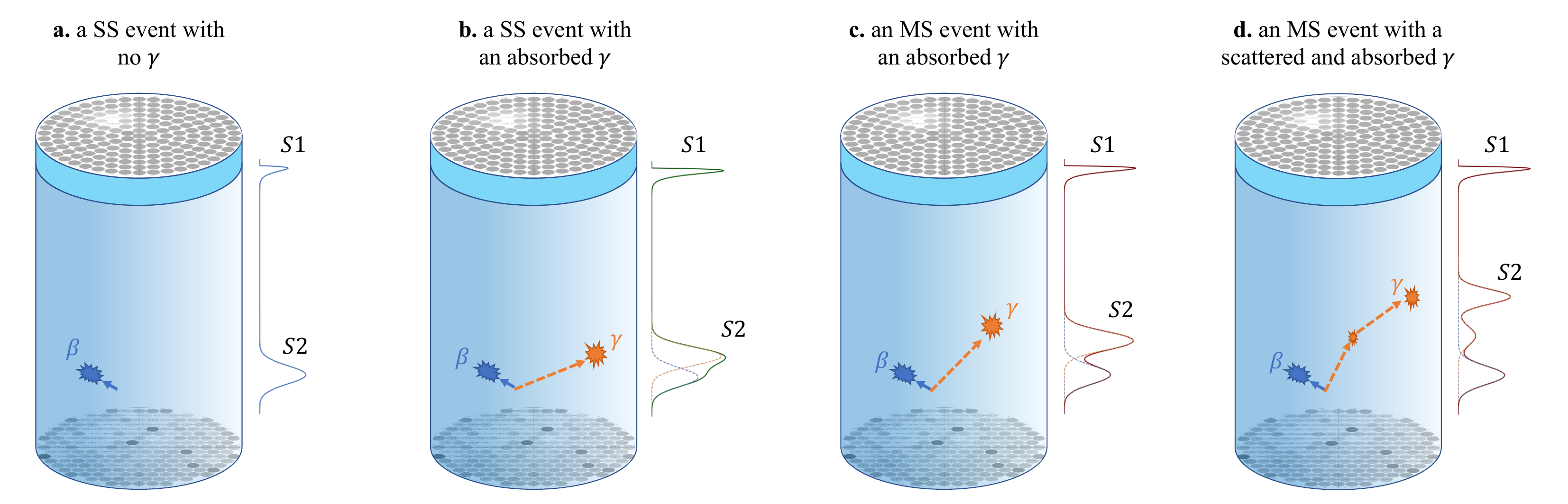}
    \caption{Schematic diagrams of tomographic signatures of GS and ES decays. 
    a. a $\beta$ from GS decay registers one single \emph{S2} and is identified as a SS event. 
    b. $\beta$ and $\gamma$ from an ES decay register overlapping \emph{S2}s and are identified as a SS event. 
    c. $\beta$ and $\gamma$ from an ES decay register separated \emph{S2}s and are identified as an MS event. 
    d. a $\gamma$ from an ES decay may experience one or more scattering before been absorbed in LXe. The $\beta$ and $\gamma$ vertices register separated \emph{S2}s and are identified as an MS event.
    }
    \label{fig:detector_response_S2}
\end{figure}

The PandaX-4T detector is a cylindrical dual-phase liquid-xenon TPC with an active volume (AV) of 118.5~cm in both diameter and height, enclosed by a side electric field cage.
The AV contains 3.7~tonnes of natural xenon. 
Two photomultiplier tube (PMT) arrays, located at the top and bottom of the AV, record scintillation and ionization signals.
A detailed description of the detector is provided in Ref.~\cite{PandaX-4T:2021bab}.

The TPC reconstructs the three-dimensional position and total energy of each event.
An energy deposition produces a prompt scintillation signal (\emph{S1}) and ionization electrons, which drift upward and generate a delayed electroluminescence signal (\emph{S2}) at the liquid–gas interface.
Both \emph{S1} and \emph{S2} are recorded by the PMT arrays.
The horizontal $(x,y)$ position is reconstructed from the photon distribution on the top PMT array, while the vertical $(z)$ coordinate is derived from the time delay between \emph{S1} and \emph{S2}.
The event energy is obtained by combining the calibrated amplitudes of the two signals.
Energy calibration is performed using external $\gamma$ sources, including $^{137}$Cs, $^{60}$Co, and $^{232}$Th, while position-dependent corrections are derived from the internal source $^\mathrm{83m}$Kr.

In this work, the PandaX-4T TPC functions as a tomographic $\beta$–$\gamma$ spectrometer for measuring the BRs of $^{214}$Pb decay.
For ES decays of $^{214}$Pb occurring within the AV, both the emitted $\beta$ and accompanying $\gamma$/ICEs particles are detected.
Most $\gamma$ rays are fully absorbed in liquid xenon, as the attenuation length of the highest-energy (839~keV) line is about 6~cm—roughly one-twentieth of the TPC diameter.
Consequently, the measured ES energy spectrum begins at the full energy of the de-excitation $\gamma$ rays (Fig.~\ref{fig:Branching_ratio_database}d); for example, the $\mathrm{ES_{295}}$ $\beta$ spectrum is shifted upward by 295~keV.
The shifted ES spectra overlay the GS spectrum with relative intensities governed by their BRs.
The strong self-shielding of xenon further suppresses external $\gamma$ backgrounds in the central fiducial volume (FV).

Events containing both $\beta$ and $\gamma$/ICE energy deposits are classified as single-site (SS) or multiple-site (MS) according to the number of distinct peaks in the \emph{S2} waveform.
Figure~\ref{fig:detector_response_S2} illustrates representative SS and MS scenarios.
A $\beta$ particle from $^{214}$Pb typically produces a single localized energy deposition, resulting in one \emph{S2} peak (Fig.~\ref{fig:detector_response_S2}a).
For ES decays, the emitted $\gamma$ ray may undergo multiple Compton scatterings before being absorbed; depending on the vertical separation between the $\beta$ and $\gamma$ vertices, the resulting $\beta$–$\gamma$ cascade can appear as either an SS or MS event (Fig.~\ref{fig:detector_response_S2}).
The ability to discriminate between SS and MS events enables the PandaX-4T TPC to further isolate energy deposits from individual decay branches of $^{214}$Pb.
\begin{figure}[t]
    \centering
    \includegraphics[width=0.6\columnwidth]{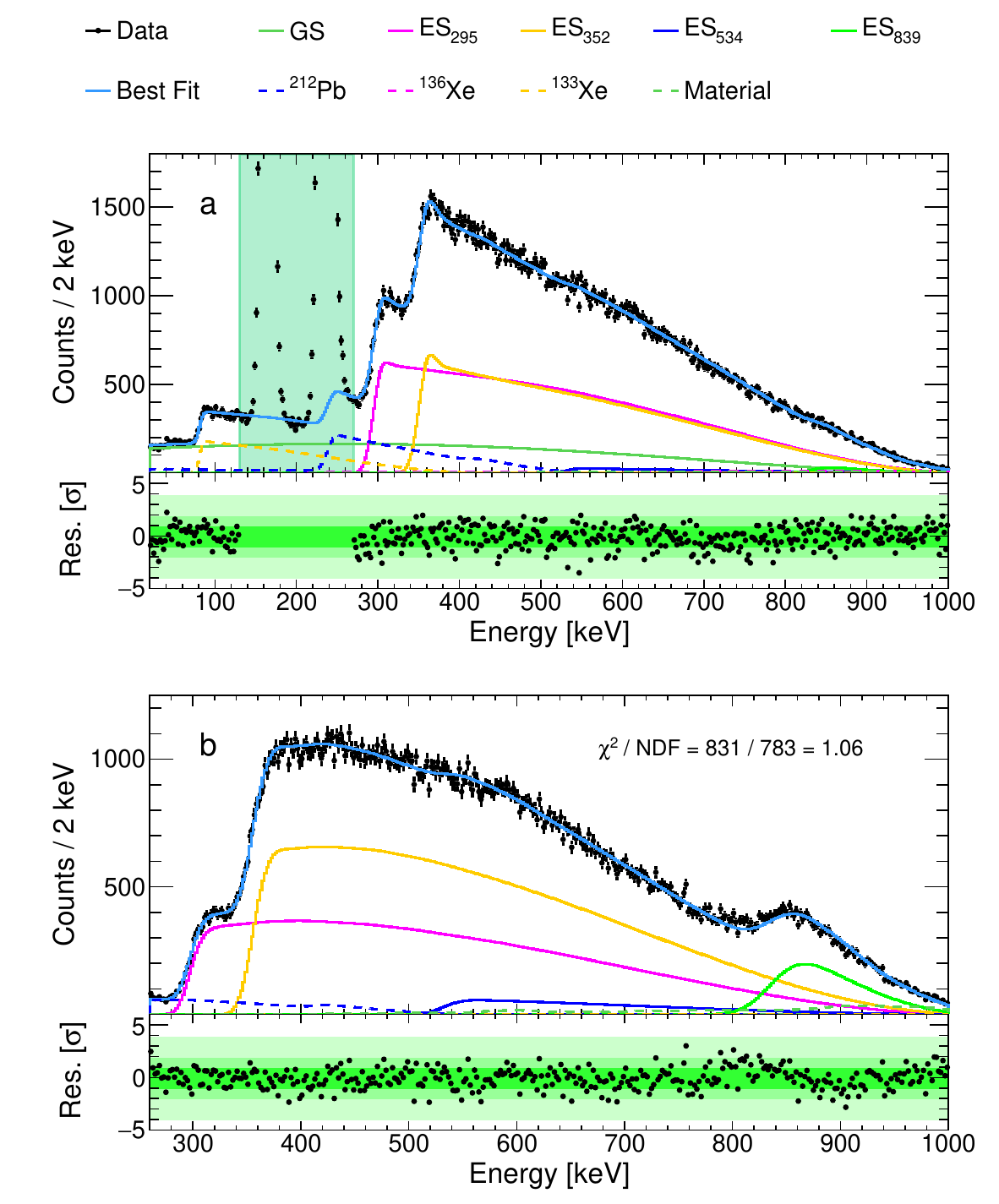}
    \caption{The $^{214}$Pb data and the final best-fit are shown for SS (a) and MS (b), with a bin size of 2 keV. 
    The horizontal axis represents the reconstructed energy in the data. The shaded green area in (a) represents the excluded region.
    The lower panel in each figure shows the fit residuals together with $\pm 1 \sigma$, $\pm 2 \sigma$, and $\pm 4 \sigma$ bands. }    
    \label{fig:fitted_output}
\end{figure}

\section{$^{214}$Pb data and fitted BRs}
High-quality $^{214}$Pb $\beta$ decay data were collected during a dedicated $^{222}$Rn injection campaign.
Throughout the campaign, the average $^{214}$Pb activity in the FV was approximately 0.5~Bq.
In total, we accumulated about half a million $^{214}$Pb decay events with a signal-to-background ratio exceeding 10 after all selection procedures.

Data production and event selection follow procedures established in previous PandaX-4T publications~\cite{PandaX:2022kwg, PandaX:2023ggs}.
We utilize the SS spectrum from 20 to 1000~keV (Fig.~\ref{fig:fitted_output}a) and the MS spectrum from 260 to 1000~keV (Fig.~\ref{fig:fitted_output}b), selecting events from the central, cleanest $1118 \pm 24$~kg of xenon.
Two prominent monoenergetic peaks at 164~keV (from $\textsuperscript{131m}$Xe) and 236~keV (from $\textsuperscript{129m}$Xe and $^{127}$Xe) in the SS spectrum originate from prior neutron calibrations.
Details of data quality cuts and SS identification criteria are provided in the Methods section.
The SS selection criteria, applied consistently to MC simulations and external $\gamma$ source calibration data, select statistically identical SS fractions, validating the selection procedure.
Differences between MC and data in SS (MS) fractions within the region of interest (ROI) average 3.42\% (3.31\%), and these values are conservatively assigned as systematic uncertainties for all signal and background components.

\begin{figure}[t]
\centering
\includegraphics[width=0.6\columnwidth]{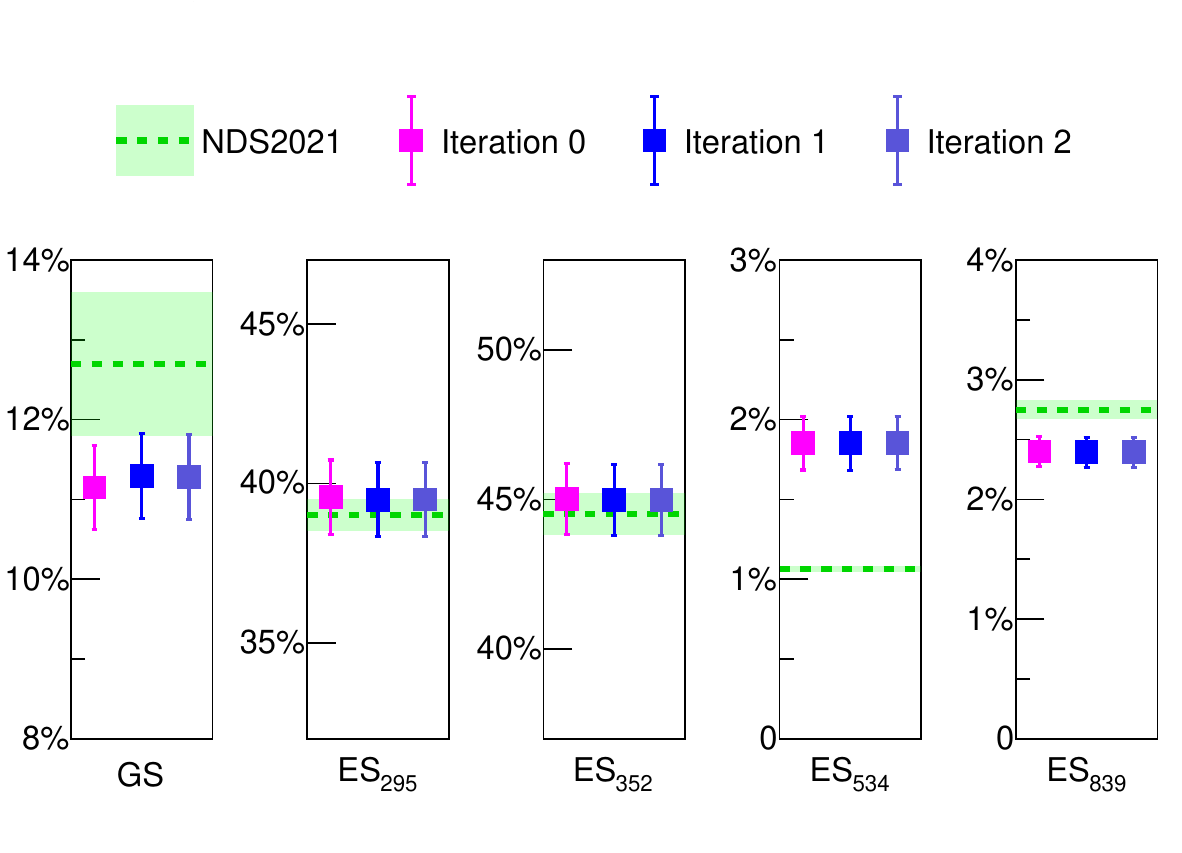}
\caption{Fitted branching ratios of $^{214}$Pb $\beta$ decay for each iteration.
The solid squares represent the best-fit results with $1\sigma$ error bars.
The NDS2021 values are shown as the green dashed line with error bands.}
\label{fig:fitted_result}
\end{figure}

Theoretical $\beta$ spectra for the five dominant $^{214}$Pb decay branches were computed and convolved with the PandaX-4T detector response.
Each $\beta$ spectrum was calculated using the state-of-the-art $\beta$ decay code~\cite{Haaranen2017}, incorporating updated atomic exchange corrections and refinements from the next-to-leading-order formalism~\cite{Hayen2018, Nitescu2023}.
Among the five branches, only the $\mathrm{ES_{839}}$ transition is allowed and therefore exhibits a universal spectral shape~\cite{Suhonen2007}.
The remaining transitions are forbidden, and their spectral shapes depend on experimental BRs and nuclear model parameters~\cite{Ramalho2024, Ramalho2024a, Ramalho2024b}.
Initial theoretical $\beta$ curves were generated using BRs from NDS2021 and subsequently passed through the full MC simulation pipeline, including energy deposition and detector response modelling.
Further details of these calculations and simulations are provided in Methods.

We performed a simultaneous binned likelihood fit of the SS and MS spectra using the simulated curves to extract the BRs of the five $^{214}$Pb decay branches.
The fit ranges are demonstrated in Fig.~\ref{fig:fitted_output} with 130–270~keV excluded.
Nuisance parameters, including detection efficiency and energy response, were constrained with Gaussian penalty terms in the likelihood function.

The best-fit BRs were used to generate updated theoretical $\beta$ spectra, which were refitted iteratively until convergence.
After three iterations, the fitted BRs stabilized.
The final fit (Fig.~\ref{fig:fitted_output}) yields a reduced $\chi^2$/NDF = 1.06.
The resulting BRs are:
$11.3\% \pm 0.5\%$ ($\mathrm{GS}$),
$39.5\% \pm 1.2\%$ ($\mathrm{ES_{295}}$),
$45.0\% \pm 1.2\%$ ($\mathrm{ES_{352}}$),
$1.9\% \pm 0.2\%$ ($\mathrm{ES_{534}}$),
and $2.4\% \pm 0.1\%$ ($\mathrm{ES_{839}}$).
The uncertainty of the $\mathrm{GS}$ BR is about half of the previously reported value, and its central value is 11\% smaller than that listed in NDS2021—an important correction for modelling low-energy backgrounds in WIMP searches.
The BRs for $\mathrm{ES_{534}}$ and $\mathrm{ES_{839}}$ differ from the NDS2021 values by 79\% (4.7$\sigma$) and 13\% (2.4$\sigma$), respectively.

The results are cross-validated using alternative fit strategies.
A fit using the SS spectrum alone produces consistent results.
Background contributions to the spectra in Fig.~\ref{fig:fitted_output} are negligible: the total counts from material $\gamma$s, $^{136}$Xe, $^{133}$Xe, and $^{212}$Pb constitute only $(7.0 \pm 0.3)\%$ of the $^{214}$Pb events.
No significant changes are observed when fixing background contributions, energy resolution, or energy offset to their nominal values.

\section{Summary and outlook}

We have introduced the Tomographic $\beta$–$\gamma$ Spectroscopy technique, which enables the simultaneous measurement of the energies of $\beta$, $\gamma$, and ICEs—a capability previously unattainable in traditional $\beta$ decay experiments.
Leveraging vertex reconstruction, a TBGS detector resolves local energy depositions within the active medium.
It directly measures decays to both the ground and excited states using a single dataset, thereby eliminating detector normalisation ambiguities while extracting the decay scheme.
For ES transitions, TBGS detects the accompanying $\gamma$ rays and ICEs with nearly 100\% efficiency, overcoming the challenges of modelling decay schemes and correcting for $\gamma$ detection efficiencies.

Using the PandaX-4T liquid-xenon TPC as a TBGS detector, we have precisely measured the BRs of $^{214}$Pb $\beta$ decay.
A dedicated campaign yielded nearly half a million $^{214}$Pb events within the ROI, with background contributions below one tenth of the signal rate.
The measurement uncertainties are well constrained thanks to the exceptional data quality and rigorous detector response characterisation of PandaX-4T.
Our measured BRs deviate significantly from the widely used NDS2021 database values.
In particular, the BR to the $\mathrm{ES_{534}}$ state is 79\% higher than the NDS2021 value, corresponding to a 4.7$\sigma$ discrepancy, while the BR to the ground state is 11\% lower—a difference that directly impacts the extrapolated $^{214}$Pb background in the WIMP search region.

We have also computed updated $\beta$ spectra for the five principal $^{214}$Pb decay branches, combining the newly measured BRs with state-of-the-art nuclear shell model (NSM) calculations.
The tuned spectral parameters agree well with theoretical predictions~\cite{Ramalho2024,Warburton1991a}, further validating our measurement.
These spectra establish a new benchmark for $^{214}$Pb background modelling in xenon detectors, enhancing the sensitivity of dark matter and neutrino experiments to the ultra rare events.

Our work demonstrates the first realization of TBGS and marks an initial step toward revealing the full potential of liquid-xenon TPCs as precision spectrometers.
Future dedicated setups—with smaller active masses and improved energy and position resolutions—could extend TBGS to a wider range of isotopes, introduced either as particulates or through novel doping techniques.
With more advanced reconstruction algorithms, the individual $\beta$ and $\gamma$ energies in multi-site events could be resolved, further reducing systematic uncertainties.

More broadly, TBGS may provide new insight in nuclear structure studies and offers a direct experimental handle on physics beyond the Standard Model through high precision $\beta$ decay spectroscopy.
A liquid xenon TPC could ultimately serve as a versatile tomographic spectrometer for cascades of $\gamma$ rays, neutrons, $\alpha$ particles, and other charged particles of nuclear decays—enabling a broad range of applications in both fundamental and applied nuclear science.

\newpage
\section{Methods}
\subsection{Signal model of the $\beta$ spectra}

The $\beta$ spectra of $^{214}$Pb were modelled using large-scale NSM and the state-of-the-art $\beta$ decay code~\cite{Haaranen2017}, taking into account of the allowed and forbidden nature of individual decays.
The nuclear wave functions of the initial $0^+$ ground state of $^{214}$Pb and those of the ground and excited states in the daughter nucleus $^{214}$Bi were computed within the NSM  using the code KSHELL~\cite{Shimizu2019} and the \textit{khpe} Hamiltonian~\cite{Warburton1991}, as implemented in NuShellX{$@$}MSU~\cite{Brown2014}. The adopted model space included the $1h_{9/2}$, $2f_{7/2}$, $2f_{5/2}$, $3p_{3/2}$, $3p_{1/2}$, and $1i_{13/2}$ orbitals for protons, and the $1i_{11/2}$, $2g_{9/2}$, $2g_{7/2}$, $3d_{5/2}$, $3d_{3/2}$, $4s_{1/2}$, and $1j_{15/2}$ orbitals for neutrons. No configuration truncations were applied.

The $\beta$ spectra for individual transitions were calculated using the $\beta$ decay code of Ref.~\cite{Haaranen2017}, incorporating updated atomic exchange corrections~\cite{Nitescu2023}. 
This code, originally developed for allowed transitions, was employed here as a surrogate for forbidden non-unique decays as well as for the allowed branch, with other refinements of the next-to-leading-order $\beta$ formalism following Ref.~\cite{Hayen2018}. For each transition, the vector-type small relativistic nuclear matrix element (sNME)~\cite{Behrens1982} was fitted to reproduce the experimental partial half-life. This fitting effectively accounts for contributions from orbitals outside the defined model space (the frozen core and orbitals above the model space), a limitation of the NSM. 
As detailed in Refs.~\cite{Ramalho2024,Ramalho2024a,Ramalho2024b}, this procedure yields two consistent solutions for each transition: one with sNME values smaller and one with values larger than the conserved vector current (CVC) value~\cite{Behrens1982,Ramalho2024}.

In our calculations, the lower set of sNME values lies below the CVC limit, whereas the upper set exceeds it. These sNMEs contribute to the decays to the $1^-$ ground state and the 295~keV and 534~keV excited $1^-$ states. The transition to the 352~keV excited $0^-$ state does not involve an sNME, since a $J_i = 0 \to J_f = 0$ decay admits no vector component. 
Instead, a mesonic enhancement factor, $\varepsilon_{\rm MEC}$, appears in this $\Delta J = 0$ transition, whose value—taken from Ref.~\cite{Kostensalo2018}—was used together with $g_{\rm A}$ to fix the corresponding BR. 
The decay to the 839~keV $1^+$ state is an allowed transition with a universal spectral shape~\cite{Suhonen2007}. 
All transitions involve the weak axial coupling $g_{\rm A}$, while those to $1^-$ states also depend on the weak vector coupling $g_{\rm V}$, taken here as $g_{\rm V} = 1.0$ from the CVC hypothesis. 
Owing to the well-known uncertainty in the effective quenching of $g_{\rm A}$, we performed the spectral computations for each branch over a range of $g_{\rm A}$ values from 0.6 to 1.2 in steps of 0.1.

\begin{figure}[t]
 \centering
    \includegraphics[width=0.6\columnwidth]{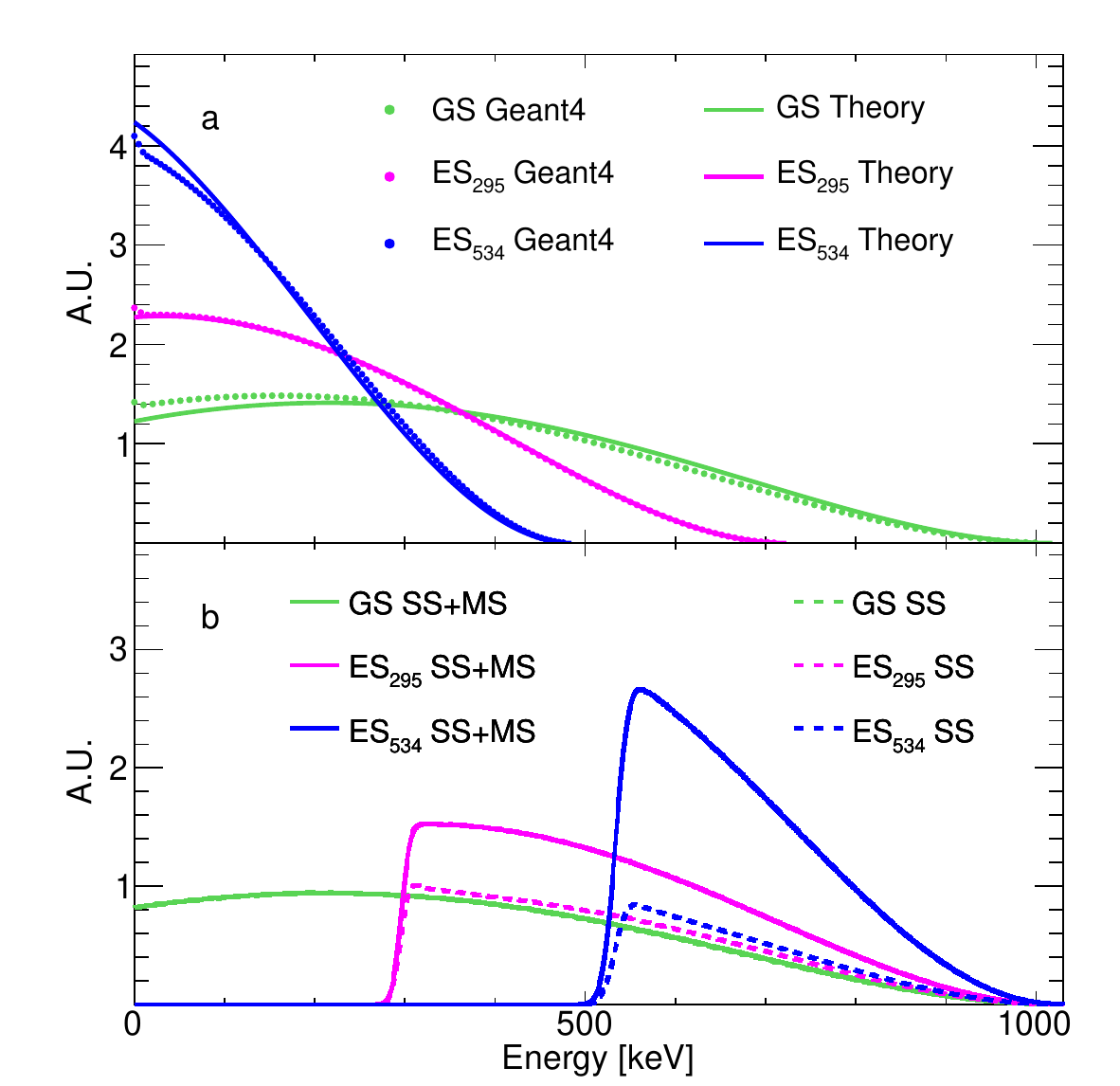}
    \caption{Comparison of the energy response spectra of the detector due to the decay branches of $\mathrm{GS}$, $\mathrm{ES_{295}}$, and $\mathrm{ES_{534}}$ using Geant4 default $\beta$ energy spectra and using theoretical $\beta$ energy spectra. The top graph is the $\beta$ energy spectra, and the bottom is the detector response energy spectra.
    }
    \label{fig:detect_response_spectra}
\end{figure}

Fig.~\ref{fig:detect_response_spectra}a shows the comparison of $\mathrm{GS}$, $\mathrm{ES_{295}}$ and $\mathrm{ES_{534}}$ spectra of the latest theoretical calculation and the Geant4 model.
Fig.~\ref{fig:detect_response_spectra}b takes into account the de-excitation $\gamma$-ray(s), and the expected spectra are shifted by the excitation energy in the PandaX-4T detector.
The curves included energy resolution and SS/MS discrimination for both the theoretical calculation and Geant4.
Since the mean free path of the 534~keV $\gamma$ is relatively long, more events are tagged as MS than SS. 

The theoretical curves are grouped as lower and upper sets.
The lower set has $\mathrm{GS}$ and $\mathrm{ES_{295}}$ spectra generated with smaller sNME values, the $\mathrm{ES_{534}}$ spectrum with a larger sNME value, whilst the upper set has all three spectra with larger sNME values. 
Although the spectra were computed for $g_{\rm A}$ values from 0.6 to 1.2, a value of $g_{\rm A}=1.0$ was adopted for this work. 
This choice is informed by the extensive study of first-forbidden $\beta$ decay transitions in the lead region by Warburton~\cite{Warburton1991a}, which suggests a quenched value of $g_{\rm A}\approx 1.0$. 
For transitions where $\Delta J=0$, both $g_{\rm A}$ and the mesonic enhancement factor, $\varepsilon_{\rm MEC}$, contribute. 
With $g_{\rm A}$ fixed at 1.0, $\varepsilon_{\rm MEC}$ was adjusted to reproduce the experimental BRs. 
This procedure resulted $\varepsilon_{\rm MEC}$ values of 2.117 and 2.153 for the lower and upper sets, respectively, for the last iteration, consistent with Warburton's~\cite{Warburton1991a} value 2.01. 
These results are also consistent with our previous findings where the value $\varepsilon_{\rm MEC}\approx 2.2$ for $g_{\rm A}=1.0$ was recorded in~\cite{Ramalho2024}. 
The values are slightly reduced from those found in the earlier work because the present calculations now incorporate more spectral corrections, such as improved atomic exchange corrections relevant for low electron energies. 

\subsection{$^{214}$Pb data in PandaX-4T}

The $^{214}$Pb data production and event selection follow procedures consistent with previous PandaX-4T analyses~\cite{PandaX:2022kwg, PandaX:2023ggs}.
For the analysis, we use the SS spectrum from 20~keV to 1~MeV and the MS spectrum from 260~keV to 1~MeV.
Each event position is reconstructed from the \emph{S2} charge pattern on the top PMT array.
For MS events, the total \emph{S1} charge, $Q_{S1}$, is partitioned into $Q_{S1_i}$ according to the corresponding charge fraction of each \emph{S2} signal ($S2_i$, where $i = 1, 2, \dots$).
Each $Q_{S1_i}$ is corrected for position dependence and summed to obtain $Q_{S1}^{\mathit{c}}$.
Similarly, the \emph{S2} charge on the bottom PMT array, $Q_{S2_{bi}}$, is corrected for electron drift lifetime and mapped according to external $\gamma$ calibration sources to yield the summed charge $Q_{S2_{b}}^{\mathrm{c}}$.
The event energy is then reconstructed from the ($Q_{S1}^{\mathit{c}}$, $Q_{S2_{b}}^{\mathrm{c}}$) pair.
The top-array \emph{S2} charge is excluded from energy reconstruction due to PMT saturation at the MeV scale~\cite{PandaX:2022kwg}.

The $^{214}$Pb decay events are selected within the central FV defined by $R^2 \in [0, 2000~\mathrm{cm^2}]$ and $Z \in [-307, 307]$~mm, with the origin set at the geometric center of the liquid xenon volume.
The FV corresponds to a xenon mass of $1118 \pm 24$~kg, determined from the fraction of uniformly distributed $^{83\mathrm{m}}\mathrm{Kr}$ events relative to the full AV~\cite{PandaX:2023ggs}.
Events overlapping in time, defined as those with an additional \emph{S1} signal within 1~ms before or during the event window, are removed.
$\beta$ particles from $^{214}$Bi and $^{212}$Bi decays are rejected via coincidence with subsequent $\alpha$ decays from $^{214}$Po and $^{212}$Po, respectively, which have half-lives of 164.3~$\mu$s and 0.3~$\mu$s.
Using a 2.5~ms coincidence window preceding the $\alpha$ \emph{S1} signal, we eliminate 99.9974\% (100\%) of $^{214}$Bi ($^{212}$Bi) $\beta$ events.

Data quality cuts and SS identification criteria are identical to those used in Ref.~\cite{PandaX:2023ggs}.
The efficiency of these cuts, which remove noise and select electron recoil events, is $99.87 \pm 0.03$\% for SS and $99.86 \pm 0.03$\% for MS.
SS/MS discrimination is modelled using BambooMC, a Geant4-based simulation framework developed by PandaX~\cite{Chen:2021asx}.
Simulated SS and MS fractions for both signal and background events within the ROI are obtained by converting Geant4 energy depositions into individual \emph{S2} signals, smeared in time according to measured drift-time-dependent diffusion.
The resulting pseudo-\emph{S2} waveforms are processed through the same SS/MS discrimination algorithm as the data.
The comparison between simulation and calibration data shows no energy-dependent bias; the average fractional differences, 3.42\% (SS) and 3.31\% (MS), are conservatively taken as systematic uncertainties for all components.

Backgrounds within the ROI arise from intrinsic radioactivity in detector materials and xenon.
$\gamma$ rays from $^{60}$Co, $^{40}$K, $^{232}$Th, and $^{238}$U in surrounding components are significantly attenuated by xenon self-shielding and are grouped following Ref.~\cite{PandaX:2022kwg}.
The contribution from double $\beta$ decay of $^{136}$Xe is constrained using its measured half-life in PandaX-4T~\cite{PandaX:2022kwg}.
A small $^{220}$Rn contamination is introduced during the $^{222}$Rn injection, and $^{133}$Xe is generated via neutron activation during calibration.
The activities of $^{212}$Pb and $^{133}$Xe are treated as free parameters in the fit.
Two monoenergetic peaks at 164~keV (from $\textsuperscript{131m}$Xe) and 236~keV (from $\textsuperscript{129m}$Xe and $^{127}$Xe), both originating from previous neutron calibrations, are excluded from the fit.

\subsection{Fit and results}

\begin{table}[b]
    \caption{
    The total background contributions in the ROI (SS + MS). 
    The fitted counts are obtained from the iteration 2 lower and iteration 2 upper $\beta$ spectra fitting.}
    \label{tab:bkg_summary}
    \centering
    \renewcommand{\arraystretch}{1.3}
    \begin{tabular}{cccc}
    \toprule
    \multicolumn{1}{c}{Components} & \multicolumn{1}{c}{Expected} & \multicolumn{1}{c}{Fitted} & \multicolumn{1}{c}{Fitted} \\
     & & (lower) & (upper) \\
    \midrule
    Material & 5206 $\pm$ 223 & 6008 $\pm$ 254 & 6002 $\pm$ 250 \\
    $^{136}$Xe & 2392 $\pm$ 110 & 2556 $\pm$ 132 & 2799 $\pm$ 139 \\
    $^{212}$Pb & float & 17777 $\pm$ 841 & 18503 $\pm$ 845 \\
    $^{133}$Xe & float & 6110 $\pm$ 308 & 6600 $\pm$ 316 \\
    \midrule
    Overall efficiency (SS) & 3.4\% & 2.7\% $\pm$ 3.1\% & 8.3\% $\pm$ 3.1\% \\
    Overall efficiency (MS) & 3.3\% & 5.7\% $\pm$ 3.2\% & 3.0\% $\pm$ 3.0\% \\
     \bottomrule
    \end{tabular}
\end{table}

\begin{table}[b]
  \caption{
    Summary of sources of systematic uncertainties. 
    $\mathcal{M}_0$ represents the 5-parameter detector response model (see text), with means and uncertainties determined from calibration fit.
  }
  \label{tab:sys_err}
  \centering
  \renewcommand{\arraystretch}{1.5}
  \begin{tabular}{>{\centering\arraybackslash}m{2.5cm}>{\centering\arraybackslash}m{2.1cm}>{\centering\arraybackslash}m{2.5cm}}
    \toprule
    \multicolumn{2}{c}{Sources} & \multicolumn{1}{c}{Values} \\
    \midrule
    \multirow{5}{*}{$\mathcal{M}_0$ (SS)} 
      & $a_0$~[$\sqrt{\mathrm{keV}}$] & $0.48\pm0.03$ \\
      & $b_0$~[keV$^{-1}$] & $(6.1\pm1.0) \times 10^{-6}$ \\
      & $c_0$ & $(-2.3\pm2.2) \times 10^{-3}$ \\
      \cmidrule(r){2-3}
      & $d_0$ & $0.9984\pm0.0004$ \\
      & $e_0$~[keV] & $0.72\pm0.56$ \\
      \cmidrule(r){1-3}
      \multirow{5}{*}{$\mathcal{M}_0$ (MS)} 
      & $a_0$~[$\sqrt{\mathrm{keV}}$] & $0.41\pm0.18$ \\
      & $b_0$~[keV$^{-1}$] & $(14.0\pm2.1) \times 10^{-6}$ \\
      & $c_0$ & $(-3.7\pm8.6) \times 10^{-3}$ \\
      \cmidrule(r){2-3}
      & $d_0$ & $0.9995\pm0.0005$ \\
      & $e_0$~[keV] & $0.66\pm0.87$ \\
      \cmidrule(r){1-3}
      \multirow{2}{*}{Overall efficiency (SS)} 
      & $^{232}$Th SS fraction & ($49.2\pm3.4$)\% \\
      & Quality cut & ($99.87\pm0.03$)\% \\
      \cmidrule(r){1-3}
      \multirow{2}{*}{Overall efficiency (MS)} 
      & $^{232}$Th MS fraction & ($57.8\pm3.3$)\% \\
      & Quality cut & ($99.86\pm0.03$)\% \\
      \cmidrule(r){1-3}
      \multicolumn{2}{c}{Background model} & \multicolumn{1}{c}{Table~\ref{tab:bkg_summary}} \\
      \bottomrule
  \end{tabular}
\end{table}

A simultaneous binned-likelihood fit of the SS and MS spectra is performed, with the BRs of the five dominant $^{214}$Pb decay branches left free.
The SS spectrum covers 20–1000~keV (excluding 130–270~keV), while the MS spectrum starts from 260~keV.
The likelihood function is defined as:
\begin{equation}
 L =  \prod_{r=0}^{1} \prod_{i=1}^{N_\mathrm{bins}} \frac{(N_{r,i})^{N_{r,i}^\mathrm{obs}}e^{-N_{r,i}}}{N_{r,i}^\mathrm{obs}!}
 \mathcal{G}(\mathcal{M}_r; \mathcal{M}_r^0, \Sigma_r)  \cdot \prod_{j=1}^{N_\mathrm{G}} G(\eta_j; 0, \sigma_j),
\label{eq::likelihood}
\end{equation},
where $N_{r,i}$ and $N_{r,i}^{\mathrm{obs}}$ denote the expected and observed counts in the $i$-th energy bin for SS or MS spectra ($r = 0, 1$).
The Gaussian penalty term $\mathcal{G}(\mathcal{M}; \mathcal{M}_0, \Sigma_r)$ of the energy response contains the five-parameter $\mathcal{M}_0$ and the covariant matrix $\Sigma_r$.
The Gaussian penalty terms $G(\eta_j; 0, \sigma_j)$ constrain the nuisance parameters $\eta_{a}$ and $\eta_{b}$, which are the relative uncertainties of the overall efficiency and the background components, respectively.
$N_\mathrm{G}=14$ is the number of Gaussian-constrained nuisance parameters.
For each spectrum, the expected bin count is given by
\begin{equation}
 N_i =  (1+\eta_{a}) \cdot [ \sum_{s=1}^{N_\mathrm{sig}} n_{s} \cdot S_{s, i}
  + \sum_{b=1}^{N_\mathrm{bkg}} (1+\eta_b) \cdot n_{b} \cdot B_{b, i}],
\end{equation},
where $n_s$ and $n_b$ are the total counts of signal $s$ and background component $b$, respectively.
$S_{s,i}$ and $B_{b,i}$ are normalized spectra of signal and background components.

Theoretical $\beta$ spectra are convolved with the PandaX-4T detector response and iteratively fitted to the data.
Two initial theoretical sets—denoted as “lower” and “upper” according to distinct sNME choices—are computed using BRs from NDS2021.
Updated theoretical spectra are regenerated after each iteration using the fitted BRs until convergence is achieved (after three iterations).
The lower sNME set provides a significantly better fit, with $\Delta\chi^2 = 48$ compared to the upper set in Iteration 0.
The final (Iteration 2) fit yields a reduced $\chi^2/\mathrm{NDF} = 1.06$ and the following BRs:
$11.3\% \pm 0.5\%$ ($\mathrm{GS}$), $39.5\% \pm 1.2\%$ ($\mathrm{ES_{295}}$), $45.0\% \pm 1.2\%$ ($\mathrm{ES_{352}}$), $1.9\% \pm 0.2\%$ ($\mathrm{ES_{534}}$), and $2.4\% \pm 0.1\%$ ($\mathrm{ES_{839}}$.
The BRs to $\mathrm{ES_{534}}$ and $\mathrm{ES_{839}}$ differ from the NDS2021 values by 4.7$\sigma$ and 2.4$\sigma$, respectively, while the GS BR is 11\% lower—an important shift for background modelling in the WIMP ROI.

The contributions of other background components, summarized in Table~\ref{tab:bkg_summary}, are consistent with their expected values, except for material and overall efficiency (MS), which are slightly pulled upward by 2.3 $\sigma$ and 1.7 $\sigma$.

The energy response is modelled with five parameters (Table~\ref{tab:sys_err}). 
The energy resolution is modelled as a Gaussian function with the width $\sigma(E)$ constructed as $\frac{\sigma(E)}{E}=\frac{a}{\sqrt{E}} + b \cdot E + c$, with energy in the unit of keV.
The energy scale is defined as $E = d \cdot \hat{E} + e$ to account for possible bias with respect to the reconstructed energy $\hat{E}$.
The measured energy spectrum is a convolution of the true energy spectrum with the five-parameter response model. 
The parameters and their uncertainties are determined by fitting the peaks of 41.5~keV~(from $^\textrm{83m}$Kr), 164~keV~(from $^\textrm{131m}$Xe), 236~keV~(from $^\textrm{127}$Xe and $^\textrm{129m}$Xe) obtained during calibration runs, and the 1460 keV peak (from $^\textrm{40}$K), 1764 keV peak (from $^\textrm{238}$U chain), 2615 keV peak (from $^\textrm{232}$Th chain) outside the ROI. 
These calibration peaks are completely uncorrelated with those used in the final spectral fit.
The extracted values $\mathcal{M}_0={(a_0, b_0, c_0, d_0, e_0)^T}$ and uncertainties of the parameters are used as priors, together with the $5 \times 5$ covariance matrix $\Sigma_r$ in fitting the $^{214}$Pb spectra.

\section*{Competing interests}
The authors declare no competing interests.

\section*{Acknowledgements}
The PandaX project is supported in part by grants from National Key R\&D Program of China (Nos. 2023YFA1606200, 2023YFA1606202), National Science Foundation of China (Nos. 12090060, 12090062, U23B2070), and by Office of Science and Technology, Shanghai Municipal Government (grant Nos. 21TQ1400218, 22JC1410100, 23JC1410200, ZJ2023-ZD-003). PandaX collaboration thanks for the support by the Fundamental Research Funds for the Central Universities. We also thank the sponsorship from the Chinese Academy of Sciences Center for Excellence in Particle Physics (CCEPP), Thomas and Linda Lau Family Foundation, New Cornerstone Science Foundation, Tencent Foundation in China, and Yangyang Development Fund. 
M. Ramalho acknowledges support by the Oskar Huttunen Foundation, grants of computer capacity from the Finnish Grid and Cloud Infrastructure (persistent identifier urn:nbn: fi:research-infras-2016072533), and the support by CSC–IT Center for Science, Finland, for generous computational resources.
J. Suhonen acknowledges support from the NEPTUN project (PNRR-I8/C9-CF264, Contract No. 760100/23.05.2023 of the Romanian Ministry of Research, Innovation and Digitization).
Finally, we thank the CJPL administration and the Yalong River Hydropower Development Company Ltd. for indispensable logistical support and other help. 

\section*{Author contributions}

Ke Han conceived the project.
Zhe Yuan analysed data with assistance from the PandaX collaboration members.
Marlom Ramalho and Jouni Suhonen performed theoretical calculations.
Ke Han, Jianglai Liu, Zhe Yuan drafted the manuscript with assistance from Marlom Ramalho and Jouni Suhonen.
The PandaX collaboration conducted detector setup, instrumentation, and data collection.
All authors discussed the results and contributed to the paper.

\section*{Materials \& Correspondence}
Correspondence and requests for materials should be addressed to ke.han@sjtu.edu.cn (Ke Han).
\end{document}